\documentclass[mathpazo]{cicp}

\begin{document}
\title{Optimization Approach in Axisymmetric Problems of Manipulating DC Currents}

\author[Alekseev G V  et~al.]{Gennady Alekseev\affil{1}\comma\corrauth, Dmitry Tereshko\affil{1},  and Yury Shestopalov\affil{2}}
\address{\affilnum{1}\ Institute of Applied Mathematics FEB RAS,
Vladivostok 690041, Russia \\ 
\affilnum{2}\  University of G\"{a}vle, SE-801 76 G\"{a}vle, Sweden}
\emails{{\tt alekseev@iam.dvo.ru} (G.~V.~Alekseev), {\tt ter@iam.dvo.ru} (D.~A.~Tereshko), {\tt Yury.Shestopalov@hig.se} (Yu.~V.~Shestopalov)}


\begin{abstract}
Inverse problems  of electric conductivity are studied  that arise in the design of spherical shielding or cloaking shells and other functional devices used to control DC electric fields. The shells are considered consisting of a finite number of layers 
filled with homogeneous isotropic or anisotropic medium.
The inverse problems under study are reduced to control problems with 
the layer electric conductivities taken as controls. 
A numerical algorithm to solve these problems is based on particle swarm optimization.
Various results of numerical experiments are discussed. 
The findings obtained in this study  describe a broad set of specific easy-to-manufacture structures that have the highest cloaking or shielding performance in the class of layered shells.
\end{abstract}


\maketitle


\section{Introduction}

One of the main goals of cloaking is to create conditions
such that an object irradiated by an external field becomes invisible to an observer. 
In particular, the field induced by an object may be virtually equal to that produced by an external source in the absence of the object.
In this case, a cloaked object causes no visible perturbation of the external field, thermal, electrostatic, or electromagnetic. 

Cloaking of material bodies may be provided by special  covers which change significantly the reflection and transmission of the induced field. Generally, mathematical modeling of the cloaking constitutes a severe problem which may be solved only by sophisticated  mathematical approaches and specifically created numerical  methods and codes.  Therefore, there is an urgent need to formulate 
a family of cloaking  problems
admitting closed-form solutions and work out appropriate techniques of their efficient analysis.  Such problems serve as a background for further development of the mathematical  theory of cloaking. Note an important feature of this family of cloaking  problems:
they operate with simple coordinate shapes (balls, spherical shells, parallelepipeds, cylinders) 
which are easy to manufacture and can be efficiently implemented as efficient cloaking devices. 

The purpose of this study is to propose and investigate a particular  family  
that employs layered spherical shells. Such spherical  bodies are placed in the electrostatic field and the layer conductivities are taken as control parameters that enable one to efficiently determine  cloaking conditions  on the basis of the obtained explicit solutions. The developed approach enables one to design shells fabricated of simple combinations of natural isotropic materials instead of anisotropic metamaterials with exotic properties and  accurately determine their  characteristics governed by the cloaking conditions.

 There has been a great variety of approaches and statements elaborated for the analysis of cloaking by means of different statements and models.
In 2006 the transformation optics approach was proposed in \cite{Pendry06,Leo06} to solve electromagnetic cloaking problems.  Then this method has been firstly extended to acoustic cloaking  \cite{Cum07,Chen07} and then to cloaking  from thermal, magnetic, electric, and other static fields \cite{Chen08,San11,Go12,Guen12,Yang12,Han13,Oii13}.
These findings allowed researchers to design metamaterial devices for manipulating DC currents such as invisibility cloaks, illusion devices and concentrators. It should be noted that  these devices usually adopted an analogue of transformation optics using complicated resistor networks to mimic inhomogeneous and anisotropic conductivities. Another and more general principle for DC currents manipulation based on the direct solution of electric conduction equations using Fourier method has been proposed in \cite{Han14}. However, the corresponding theory is valid under the strict limitations on the initial data which ensure the existence of an exact solution.

In this work we consider the problem of manipulating the DC currents to design cloaking or shielding devices  in the general case when the theory developed 
in  \cite{Han14} is not applicable. To solve this problem we apply an optimization method. 
This approach is based on introducing the cost functional under minimization which adequately corresponds to the inverse problem of designing a device for manipulating dc currents.  Beginning with the fundamental works of A.N. Tikhonov \cite{Ti77}, this method is widely used when studying  inverse problems of electromagnetics and heat and mass transfer.

As applied to cloaking problems, the optimization approach has initiated a new cloaking scenario known as the inverse design strategy \cite{Xu13}. 
We mention the papers \cite{Popa09,Xi09} where cloaking problems were first solved using numerical optimization (an iterative gradient method in \cite{Popa09} and a genetic algorithm in \cite{Xi09}).  Publications \cite{AlLe14,ApAn, AlLe16, Al18} are devoted to theoretical analysis of cloaking problems using optimization. 
In these works some important properties of optimal solutions to the considered problems are established depending on the choice of the cost functional and the set of controls with respect to  which the functional is minimized.
Papers \cite{Fuji2018, Fuji2018dc, PeFa17,Fach2018,Fach2018e} develop the topology and discrete material optimization for designing cloaks, concentrators and other functional devices for controlling  static physical fields. 
Letters \cite{AlLeTe17c,AlLeTe17s} are devoted to the methods for designing thermal cloaks  based on the particle swarm optimization.
In \cite{MZ17,BeKb, BeKc, CaKo,Kab12,Naka,Do17} the optimization method is used for solving the related inverse problems arising in acoustics, electromagnetics and heat conduction.

This paper is airmed at theoretical and numerical analysis of the design of functional structures in the form of spherical shells for controlling the steady-state DC electric fields. 
A main attention is paid to solving the design problems  for shielding and cloaking shells.
Keeping this goal in mind, the plan of our paper is as follows.
Based on the conditions that can be technically implemented, the desired axisymmetric shell is divided into a finite number of layers where each layer is filled with a homogeneous isotropic material. As a result, 
the manipulation problem under study is reduced to solving respective finite-dimensional control problems where constant conductivities of each layer serve as control parameters.
We propose to introduce a well-defined quantity called the measure of visibility which is connected with the cloaking efficiency by simple explicit relations.

This measure serves as a clear criterion of the cloaking quality
and play the role of a cost functional to be minimized in order to find the optimal solution 
to the cloaking problem in the considered control problem. 
Another control problem is related to finding optimal solution of the shielding shell design problem. 
To solve these control problems numerically we propose an algorithm based on particle swarm optimization. 

Using this algorithm, we will demonstrate that for a certain choice of the set on which the cost functional is minimized, a version of the bang-bang principle
is valid known  in the optimal control theory (see, e.g., \cite{Alpha99}). 
According to this principle, optimal solutions take values situated either
on the lower or upper boundary of the control set.   Therefore if to choose 
the lower and upper boundaries of the control set so that they would 
correspond to two natural materials with high contrast, we will obtain the 
highly efficient optimal solutions that can be easily implemented
in practice. This idea is applied and validated in Section 4 in the
course of a series of numerical experiments and comparison of
the obtained numerical data with the results of 
solution to the problems under study determined on the basis of 
the  alternative design strategy. 
After verifying this approach numerically we formulate simple rules 
of the design of easy-to-manufacture shielding 
or cloaking spherical shells 
providing highest performances in the class of layered shells.

\section{Statement of direct and inverse problems for DC conduction model}

We start with the statement of the general problem considered in a three-dimensional cylindrical domain $D=\{ {\bf x} \equiv (x,y,z): -z_0 < z< z_0, \;\; x^2+y^2 < c^2  \}$
with specified numbers $z_0>0$ and $c>0$ (see Fig. 1). 
Let an external electric potential $U^e$ be formed by two horizontal planes $z = \pm z_0$
with different values $u_1$ and $u_2$, and the lateral surface
$\Gamma$ of cylinder $D$ be insulated. We assume that
there is a material shell $(\Omega,\sigma)$ inside $D$. Here, $\Omega$ is a
spherical layer $a < |{\bf x}| < b\}$, $b<c$ and $\sigma$ is the electric conductivity tensor of the inhomogeneous anisotropic medium filling domain $\Omega$. We assume that the
interior $\Omega_i$: $|{\bf x}|<a$ and exterior $\Omega_e$: $|{\bf x}|>b$ of $\Omega$ are
filled with the homogeneous medium having constant electric conductivity $\sigma_i=\sigma_e=\sigma_b > 0$.

\begin{figure}[h!]
\begin{center}
\includegraphics[scale=.4]{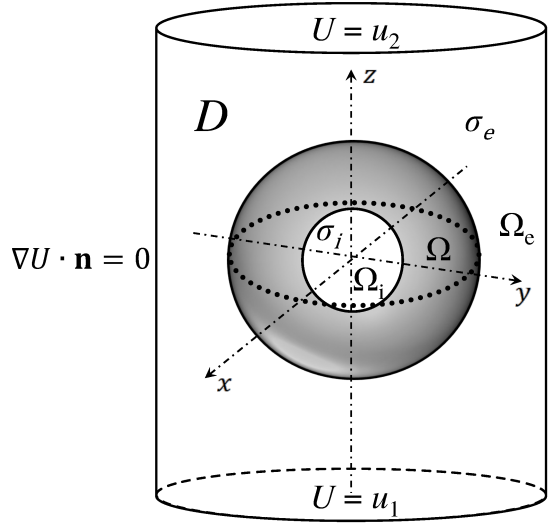}
\end{center}
\caption{Geometry of the problem.}
\label{fig:1}      
\end{figure}

In this case, the direct DC conduction problem is to determine a triple $U$ of functions: $u_i$ in $\Omega_i$,
$u$ in $\Omega$, and $u_e$ in $\Omega_e$ which satisfy the equations
\begin{equation}
\label{eq1}
\sigma_b \Delta u_i = 0 \; \mbox{ in }\; \Omega_i,
\end{equation}
\begin{equation}
{\rm div}\, (\sigma \, {\rm grad} \, u) = 0\; \mbox{ in }\; \Omega,
\end{equation}
\begin{equation}
\sigma_b \Delta u_e = 0 \; \mbox{ in } \; \Omega_e, 
\end{equation}
the boundary conditions
\begin{equation}
U|_{z= -z_0} = u_1, \;\; U|_{z= z_0} = u_2, \;\; \frac{\partial U}{\partial n}|_{\Gamma} = 0
\end{equation}
and matching conditions on boundaries $\Gamma_i$ and  $\Gamma_e$ of layer $\Omega$ in the form
\begin{equation}
u_i = u, \; \; \sigma_b \frac{\partial u_i}{\partial n} = (\sigma \nabla u) \cdot {\bf n} \mbox{ on }\Gamma_i,
\end{equation}
\begin{equation}
u_e = u, \; \; \sigma_b \frac{\partial u_e}{\partial n} = (\sigma \nabla u) \cdot {\bf n} \mbox{ on } \Gamma_e.
\label{eq6}
\end{equation}

Our goal is to analyze  and solve inverse problems for the model (\ref{eq1})--(\ref{eq6}) associated with the design of shielding or cloaking shells and other functional devices for 
controlling static DC electric fields. Generally, the inverse problems consist in finding conductivity tensor $\sigma$  of the medium filling domain $\Omega$ from the two independent conditions
\begin{equation}
\label{IP}
\nabla U =  \nabla u_i^d \; \mbox{ in } \; \Omega_i, \;\;
 U = u_e^d  \mbox{ in } \; \Omega_e. 
\end{equation}
Here, $U=(u_i,u,u_e)$ is the solution of (\ref{eq1})--(\ref{eq6}) 
while $u_i^d$  and  $u_e^d$ are given  fields in $\Omega_i$ and $\Omega_e$.
This inverse problem is often referred to as an  illusion or  camouflage  problem. 
In the particular case when $u_i^d=0$, $u_e^d=U^e$  so that (\ref{IP}) has the form 
\begin{equation}
\label{IP2}
\nabla U = 0 \mbox{ (i.e. } U=u^0) \mbox{ in } \Omega_i, \;\;
U = U^e \mbox{ in } \Omega_e
\end{equation}
where $u^0={\rm const}$, this problem is called a general cloaking problem. 
The second equality in (\ref{IP2}) means physically that the external scattered response 
$u^s_e \equiv u_e - U^e$ vanishes in the exterior $\Omega_e$ of $\Omega$. 
The shell $(\Omega, \sigma_r, \sigma_\theta)$ which ensures exact fulfillment of the conditions (\ref{IP2}) is called   a perfect cloaking shell or simply a cloak.
In the case when $\sigma_r$ and  $\sigma_\theta$ are determined solely from the first (or second) condition in (\ref{IP2})
we will refer to the corresponding problem as a shielding (or external cloaking)  problem.
When the first condition in (\ref{IP2}) is replaced by 
$\nabla U = - \nabla U^e$ in $\Omega_i$ this inverse problem is reffered to as 
the DC electric field inversion problem.

First, we will study some properties of solutions to direct problem (\ref{eq1})--(\ref{eq6}). 
In spite of the different geometries (cylindrical and spherical) of the above-introduced domains $D$ and
$\Omega \subset D$, their common important property is that they
are both axisymmetric. This feature will be essentially used below along with the following
 assumption: tensor $\sigma$ is diagonal in spherical coordinates $r$, $\theta$, $\varphi$, 
and its diagonal components (radial, polar, and azimuthal conductivities) $\sigma_r$, $\sigma_\theta$, and $\sigma_\varphi$ are independent of $\varphi$ and satisfy the assumptions 
\begin{equation}
\label{assum9}
\sigma_r > \sigma_r^0={\rm const} > 0, \;\; \sigma_\theta > \sigma_\theta^0={\rm const} > 0, \; \; \sigma_\varphi =\sigma_\theta.
\end{equation} 
Note that the condition $\sigma_\varphi =\sigma_\theta$ is widely used when
analyzing three-dimensional problems of designing spherical functional devices.

Moreover, it is well known that in the class of diagonal tensors $\sigma \equiv {\rm diag} (\sigma_r, \sigma_\theta, \sigma_\varphi)$ satisfying condition (\ref{assum9}), 
provided that $u_1 = {\rm const}$, $u_2 = {\rm const}$, there exists
an exact solution $(\sigma_r^*, \sigma_\theta^*, \sigma_\varphi^*)$ to the general cloaking problem that can be constructed using the transformation optics method. 
The solution is defined by  (see, e.g., \cite{Guen12})
\begin{equation}
\label{ES}
\sigma_r^* = \frac{b}{b-a}\left( \frac{r-a}{r} \right)^2 \sigma_b, \; 
\sigma_\theta^*=\sigma_\varphi^* = \frac{b}{b-a} \sigma_b.
\end{equation}
However, solution (\ref{ES}) has a significant drawback. 
It is not technically feasible in the sense that there are no natural anisotropic materials  with the electric conductivity properties described by formulas (\ref{ES}).

The solution (\ref{ES}) can be simplified by replacing the couple $(\sigma_r^*,\sigma_\theta^*) $ in (\ref{ES}) 
with a constant  couple  $(\sigma_r,\sigma_\theta)$ providing approximate fulfillment of conditions (\ref{IP2}). 
To find such a  couple we choose any $\sigma_r \geq \sigma^0_r$, $\sigma_\theta \geq \sigma^0_\theta$ satisfying the so-called admissibility condition
\begin{equation}
\label{AC}
2\sigma_r \sigma_\theta = \sigma^2_b + \sigma_b \sigma_r. 
\end{equation}

A simple analysis shows that for any admissible couple $(\sigma_r, \sigma_\theta)$ 
there exists an exact solution to the direct problem (\ref{eq1})--(\ref{eq6}) independent of $\varphi$. 
This solution is defined by the formulas (see, e.g., \cite{AlLeTe17s})
\[
u_i (r, \theta) =  \frac{u_0}{z_0} \left( \frac{a}{b} \right)^{q(s) -1} r\cos \theta 
+\frac{u_1 +u_2}{2} \; \mbox{ in } \; \Omega_i,  
\]
\[
u (r, \theta) =  \frac{u_0}{z_0} \left( \frac{r}{b} \right)^{q(s) -1} r\cos \theta 
+\frac{u_1+u_2}{2} \; \mbox{ in } \; \Omega,
\]
\begin{equation}
\label{eq7}
u_e (r, \theta) =  \frac{u_0}{z_0} r \cos  \theta +\frac{u_1+u_2}{2} \; \mbox{ in } \; \Omega_e 
\end{equation}
where
\[
q (s) \equiv \frac{-1+\sqrt{1+8 s} }{2}, \;\;
s=\frac{\sigma_\theta}{\sigma_r},\; \; u_0=\frac{u_2-u_1}{2}.
\]
The parameter $s={\sigma_\theta}/{\sigma_r}$ characterizes the degree of
anisotropy of the shell $(\Omega, \sigma_r, \sigma_\theta)$. 
In the particular case $s = 1$ (provided that
$\sigma_r=\sigma_\theta = \sigma_0$ so that the entire medium filling the
domain $D$ is homogeneous and isotropic), formula (\ref{eq7}) is transformed into a unified expression
\begin{equation}
U^e (z)= \frac{u_0}{z_0}r \cos  \theta +\frac{u_1+u_2}{2} = u_0 \frac{z}{z_0} +\frac{u_1+u_2}{2}.
\label{eq8}
\end{equation}
Formula (\ref{eq8}) describes the externaly applied field $U^e(z)$ used to detect objects located in domain $D$. It follows from
(\ref{eq7}) and (\ref{eq8}) that
\begin{equation}
\label{eq9}
u_e= U^e, \;\;
|\nabla u_e|= |\nabla U^e|= \frac{|u_0|}{z_0} \; \mbox{ in } \; \Omega_e, \;\;
|\nabla u_i|= \frac{|u_0|}{z_0} {\cal K} (s) \mbox{ in } \Omega_i, \;\;
{\cal K} (s) \equiv \left( \frac{a}{b} \right)^{q (s)-1}.
\end{equation}

Assuming $s\geq 1$ we define the function ${\cal M}(s)=1-{\cal K}(s)$ where ${\cal K}(s)$ is given in (\ref{eq9}). Since $a<b$ we derive from (\ref{eq9}) that $0\leq {\cal K} (s)\leq 1$ and therefore $0\leq {\cal M}(s)\leq 1$.
Moreover, we have ${\cal K} (s)\to 1$, ${\cal M} (s)\to 0$ as $s\to 1$ and
${\cal K} (s)\to 0$, ${\cal M} (s)\to 1$  as  $s\to \infty$. From (\ref{eq9}) it follows that any admissible couple $(\sigma_r,\sigma_\theta)$ ensures exact fulfillment of the second condition in 
(\ref{IP2}), whereas { the degree of fulfillment of the first condition} in (\ref{IP2}) is determined by the value  ${\cal K} (s)$ of function ${\cal K}$
 which tends to zero as $s \to \infty$. Thus, { the  larger the value of} 
$s \equiv \sigma_\theta / \sigma_r$, { the more precisely, on one hand,
	the second cloaking condition is satisfied and, on the other hand, 
the shell} $ (\Omega, \sigma_r, \sigma_\theta)$ has {  the greater degree of anisotropy} which complicates technical implementation of the corresponding cloak. 

Using the aforementioned properties of functions ${\cal M}(s)$ and ${\cal K}(s)$ and taking into account terminology of \cite{AlLeTe17s} we will refer below to ${\cal M}(s)$ and ${\cal K}(s)$ as the cloaking performance and visibility measure of the respective homogeneous cloak $(\Omega,\sigma_r,\sigma_\theta)$.  We emphasize that cloaking performance  of the homogeneous anisotropic shell 
$(\Omega,\sigma_r,\sigma_\theta)$ increases while the visibility measure decreases with the increasing anisotropy parameter $s=\sigma_\theta/ \sigma_r$.
In the limit as $s \to \infty$ we obtain the perfect cloaking shell with maximum cloaking performance ${\cal M} = 1$ and minimum visibility measure ${\cal K}=0$. 

Technical implementation of anisotropic shells is associated with great difficulties. With this in mind, our further task will be to find approximate solutions 
to the shielding or cloaking inverse problems which have the following two basic properties:

1) shielding or cloaking shells that correspond to the above solutions have a high cloaking efficiency in the sense that the corresponding conditions in (\ref{IP2}) are fulfilled with high accuracy;

2) these shields or cloaks  allow simple technical implementation in the form of a layered shell consisting of $M$ homogeneous isotropic layers filled with natural materials. 

To find these approximate solutions, we apply an optimization method. In accordance with this approach, we introduce into consideration the root-mean-square integral errors in the fulfillment of every or both conditions in (\ref{IP2}) and will find the minimums of these errors as the functions of the layers conductivities varying on some set $S$.

\section{Optimization problems for layered shell}
Let us now complicate the structure of the shell under consideration assuming that 
$(\Omega, \sigma_r$, $\sigma_\theta)$ is a layered shell consisting of $M$ concentric spherical layers
$\Omega_j$, $j = 1,2,\ldots,M$. Each of these layers is filled with a
homogeneous and (generally) anisotropic medium;
 electro-conducting properties of the layers are described by constant conductivities 
$\sigma_{rj} > 0$ and $\sigma_{\theta j} > 0$, $j = 1,2,\ldots,M$. 
The parameters $\sigma_r$ and $\sigma_\theta$ of this layered shell are given by 
\begin{equation}
\sigma_r ({\bf x}) = \sum_{j=1}^M \sigma_{rj} \chi_j ({\bf x}), \; \;
\sigma_\theta ({\bf x}) = \sum_{j=1}^M \sigma_{\theta j} \chi_j ({\bf x}).
\label{eq10}
\end{equation}
Here,  $\chi_j ({\bf x})$ is a characteristic function of layer $\Omega_j$ equal, respectively,  to $1$ and $0$ inside  and  outside $\Omega_j$.

One should emphasize that in the case of a layered shell, the inverse problems considered here become finite-dimensional because they are reduced to determination of unknown coefficients 
$\sigma_{rj}$ and $\sigma_{\theta j}$ entering formulae (\ref{eq10}). These coefficients form a $2M$-dimensional vector 
${\bf s} \equiv (\sigma_{r 1}, \sigma_{\theta 1}, \ldots , \sigma_{r M}, \sigma_{\theta M})$ having the sense of the  conductivity vector for the respective $M$-layered shell. 
In the particular case of an isotropic shell, when $\sigma_{rj}=\sigma_{\theta j}=\sigma_j$,
conductivities $\sigma_j$ of layers $\Omega_j$, $j=1,2,\ldots,M$
will form a $M$-dimensional vector ${\bf s}=(\sigma_1,\sigma_2,\ldots,\sigma_M)$.

Using an optimization method one can reduce inverse problems under study to the minimization of a certain cost functional $J$. To find the
explicit form of functional $J$, we denote by $U [{\bf s}] \equiv U(\sigma_{r 1}, \sigma_{\theta 1}, \ldots , \sigma_{r M}, \sigma_{\theta M})$ the solution to the direct problem (\ref{eq1})--(\ref{eq6})
corresponding to conductivities (\ref{eq10})  in $\Omega$ and 
to coefficient $\sigma_b$ in $\Omega_i \cup \Omega_e$. It is
assumed below that the vector ${\bf s} = (\sigma_{r 1}, \sigma_{\theta 1}, \ldots , \sigma_{r M}, \sigma_{\theta M})$
belongs to the bounded set (control set)
\begin{equation}
\label{eqS}
S= \{ {\bf s} : \, m_r  {\le} \sigma_{rj} {\le} M_r, \,  m_\theta {\le}  \sigma_{\theta j} {\le} M_\theta, \,
j=1,\ldots,M\}
\end{equation}
for specified positive constants $m_r$, $M_r$, $m_\theta$, $M_\theta$, $j=1,2, \ldots,M$. 
Let us define three cost functionals $J_i$, $J_e$, and $J$ by
\begin{equation}
\label{eq11}
J_i ({\bf s}) = \frac{\| \nabla U [{\bf s}]\|_{L^2(\Omega_i)}}{\| \nabla U^e\|_{L^2(\Omega_i)}}, \; \; 
J_e ({\bf s}) = \frac{\|U[{\bf s}]-U^e\|_{L^2(\Omega_e)}}{\|U^e\|_{L^2(\Omega_e)}}, 
\; \; 
J({\bf s})= J_e ({\bf s})+J_i ({\bf s}). 
\end{equation} 
Here, in particular, 
\begin{equation}
\label{new16a}
\|U^e\|^2_{L^2(\Omega_e)} = \int_{\Omega_e} |U^e|^2 d {\bf x},  \;\;
\|\nabla U^e\|^2_{L^2(\Omega_i)} = \int_{\Omega_i} |\nabla U^e|^2 d {\bf x}.
\end{equation}
It is clear that the value $J_i({\bf s})$ describes for any vector ${\bf s}\in S$ the mean square error of 
fulfilling the first condition in (\ref{IP2}), $J_e({\bf s})$ describes the mean square error of fulfilling the second condition in (\ref{IP2}), while $J({\bf s})$ describes the total mean square error of fulfilling both conditions in (\ref{IP2}).

Based on functionals (\ref{eq11}) we are able  now to formulate the following three control problems:
\begin{equation}
\label{eqJi}
J_i ({\bf s})   \to \min, \;\;  {\bf s}   \in S,
\end{equation} 
\begin{equation}
\label{eqJe}
J_e ({\bf s})   \to \min, \;\;  {\bf s}   \in S,
\end{equation} 
\begin{equation}
\label{eqJ}
J({\bf s})   \to \min, \;\; {\bf s}   \in S. 
\end{equation} 
We refer to problem (\ref{eqJi}) (or (\ref{eqJe})) as a shielding (or external cloaking) problem while problem (\ref{eqJ}) will be referred to as a general cloaking problem.

We emphasize that all functionals $J_i$, $J_e$, and $J$ take nonnegative values for any vector ${\bf s} \in S$.
Furthermore, it follows from (\ref{eq11}) that the condition $J_i({\bf s})=0$ is satisfied if and only if the first condition in (\ref{IP2}) is fulfilled. Similarly, the condition $J_e ({\bf s})=0$ is satisfied
if and only if the second condition in (\ref{IP2}) is fulfilled. 
Therefore problem (\ref{eqJi}) is aimed at finding an (approximate) optimal  solution of the shielding problem.
Similarly, problem (\ref{eqJe}) is aimed at finding an optimal solution of the external cloaking problem while  problem  (\ref{eqJ}) is aimed at finding an optimal solution of the general cloaking problem. 

Below we will consider two particular cases of problems  (\ref{eqJi}), (\ref{eqJe}) and (\ref{eqJ}) corresponding to different methods of choice of the set $S$ in (\ref{eqS}). 
The first case when ${\bf s} =(\sigma_r, \sigma_\theta)$ corresponds to a single-layer homogeneous anisotropic shell 
$ (\Omega, \sigma_r, \sigma_\theta)$ with constant parameters $\sigma_r$ and $\sigma_\theta$. 
The second case corresponds to a $M$-layered shell $(\Omega,{\bf s})$ consisting of isotropic layers where
${\bf s} = (\sigma_1, \sigma_2, \ldots, \sigma_M)$. 
As has been shown earlier, in the first case  when 
the additional condition $2 \sigma_r \sigma_\theta = \sigma_b^2+\sigma_b \sigma_r$ is satisfied and, besides, 
$u_1 = {\rm const}$, $ u_2 = {\rm const}$, 
there exists an exact solution to problem  (\ref{eq1})--(\ref{eq6}). It has the form
(\ref{eq7}) and satisfies (\ref{eq9}).
Using (\ref{eq9}), (\ref{eq11}) it is easy to show that for any such couple $(\sigma_r,\sigma_\theta)$ the following relations hold: 
\begin{equation}
\label{eqRav}
J_e(\sigma_r,\sigma_\theta)=0,  \;\; 
J (\sigma_r,\sigma_\theta)=J_i (\sigma_r,\sigma_\theta)={\cal K} (s).
\end{equation}
Here, $s= \sigma_\theta / \sigma_r$ and ${\cal K}(s)$ is  the  visibility measure  
of the shell $(\Omega, \sigma_r, \sigma_\theta)$.
We note that the equality $J_e(\sigma_r,\sigma_\theta)=0$ in (\ref{eqRav})
means that any mentioned couple $(\sigma_r, \sigma_\theta)$ is an exact solution to the inverse problem of external cloaking.
In addition, it follows from (\ref{eqRav}) that the value $J_i (\sigma_r,\sigma_\theta)$ of functional $J_i$, as well as the value $J(\sigma_r, \sigma_\theta)$ of functional $J$, characterizes the 
visibility measure  of the corresponding shell $(\Omega, \sigma_r,\sigma_\theta)$ which is connected with its cloaking 
performance ${\cal M}(s)$  by ${\cal M}(s)+{\cal K}(s)=1$. 
The smaller value $J(\sigma_r,\sigma_\theta)$ corresponds to smaller value of the visibility measure 
${\cal K}(s)$ which in turn corresponds to higher cloaking performance ${\cal M}(s)$
of the shell $(\Omega, \sigma_r,\sigma_\theta)$ and vice versa. 
We note that value $J({\bf s})$  will play the same role 
 in the general case of a $M$-layered shell $(\Omega,{\bf s})$.  
Therefore our goal when solving, e.g., problem (\ref{eqJ})
will consist of finding a conductivity vector (an optimal solution of (\ref{eqJ}))
${\bf s}^{opt} \in S$ for which functional $J$ takes the minimum value $J^{opt}=J({\bf s}^{opt})$
on the set $S$ and therefore the cloak $(\Omega,{\bf s}^{opt})$ possess a maximum cloaking efficiency.
We emphasize once again that mathematically $J^{opt}$ describes {the total mean square error} of fulfilling both conditions in (\ref{IP2}) achieved on the optimal solution ${\bf s}^{opt}$, 
while physically $J^{opt}$ describes the  visibility measure of the shell $(\Omega, {\bf s}^{opt})$.

To solve problems (\ref{eqJe}), (\ref{eqJ}) we use an algorithm based on the particle swarm optimization (PSO) \cite{Po07}.
Within this method, the desired parameters determining the value of minimized functional $J$ are presented
in the form of the coordinates of the position vector 
${\bf s} \equiv (\sigma_{r 1}, \sigma_{\theta 1}, \ldots , \sigma_{r M}, \sigma_{\theta M})$
of some abstract particle. A particle swarm is considered to be any finite set of particles 
${\bf s}_1, \ldots,{\bf s}_N$. Within the particle swarm optimization, one sets the initial swarm 
position ${\bf s}_0^j$, $j=1,2,\ldots,N$, and the iterative displacement procedure 
${\bf s}_j^{i+1}={\bf s}_j^i+{\bf v}_j^{i+1}$ for all particles ${\bf s}_j$, which is described by the formula (see \cite{AlLeTe17c})
\begin{equation}
\label{eqV}
{\bf v}_j^{i+1}=w {\bf v}_j^i + c_1 d_1 ({\bf p}_j^i-{\bf s}_j^i)
 + c_2 d_2 ({\bf p}_g-{\bf s}_j^i).
\end{equation}
After each displacement we calculate the value $J ({\bf s}_j^{i+1})$ of functional $J$ for the new position ${\bf s}_j^{i+1}$, compare it to the current minimum value, and, if necessary, update the individual and
global best positions ${\bf p}_j$ and ${\bf p}_g$.
In the end of this iteration process, 
all particles must come to the global minimum point.

Here, ${\bf v}_j$ is the displacement vector; $w$, $c_1$ and $c_2$ are constant parameters; and $d_1$ and $d_2$ are
random variables uniformly distributed over interval $(0,1)$. Their choice was considered in
more detail in \cite{AlLeTe17c,Po07}. The subscript $j \in \{1 , 2, \ldots , N \}$ in (\ref{eqV}) denotes the particle number and the superscript $i \in \{0 , 1, \ldots , L \}$ indicates the iteration number.

\section{Simulation results}
In this section we will discuss the results of application of the proposed optimization algorithm to solve design problems  under study for two scenarios.  {The first corresponds to the design of single-layer anisotropic shell while the second to  multilayer isotropic shells.} A main attention will be paid to a comparative analysis of the results for multilayer isotropic shells obtained using our method and a method based on the alternating design strategy.

The most time-consuming part in this algorithm is the calculation of values
$J ({\bf s}_j^i)$ of functional $J$  for the particle position ${\bf s}_j^{i+1}$
at different $i$ and $j$. This procedure  {comprises two stages}. 
At the first stage, the solution
$U[{\bf s}_j^i]$ to direct problem (\ref{eq1})--(\ref{eq6}) is calculated. 
To this end, we use the FreeFEM++ software package (www.freefem.org) designed for the numerical solution of two- and three-dimensional boundary value problems by the finite-element method.  
After determining $U[{\bf s}^i_j ]$, at the second stage, we calculate {  the mean squared integral norms} entering the definition of functionals $J_e$ or $J_i$ in (\ref{eq11}) using the formulas of numerical integration.

\begin{figure}[t!]
\centerline{
\includegraphics[width=70mm, height=60.5mm]{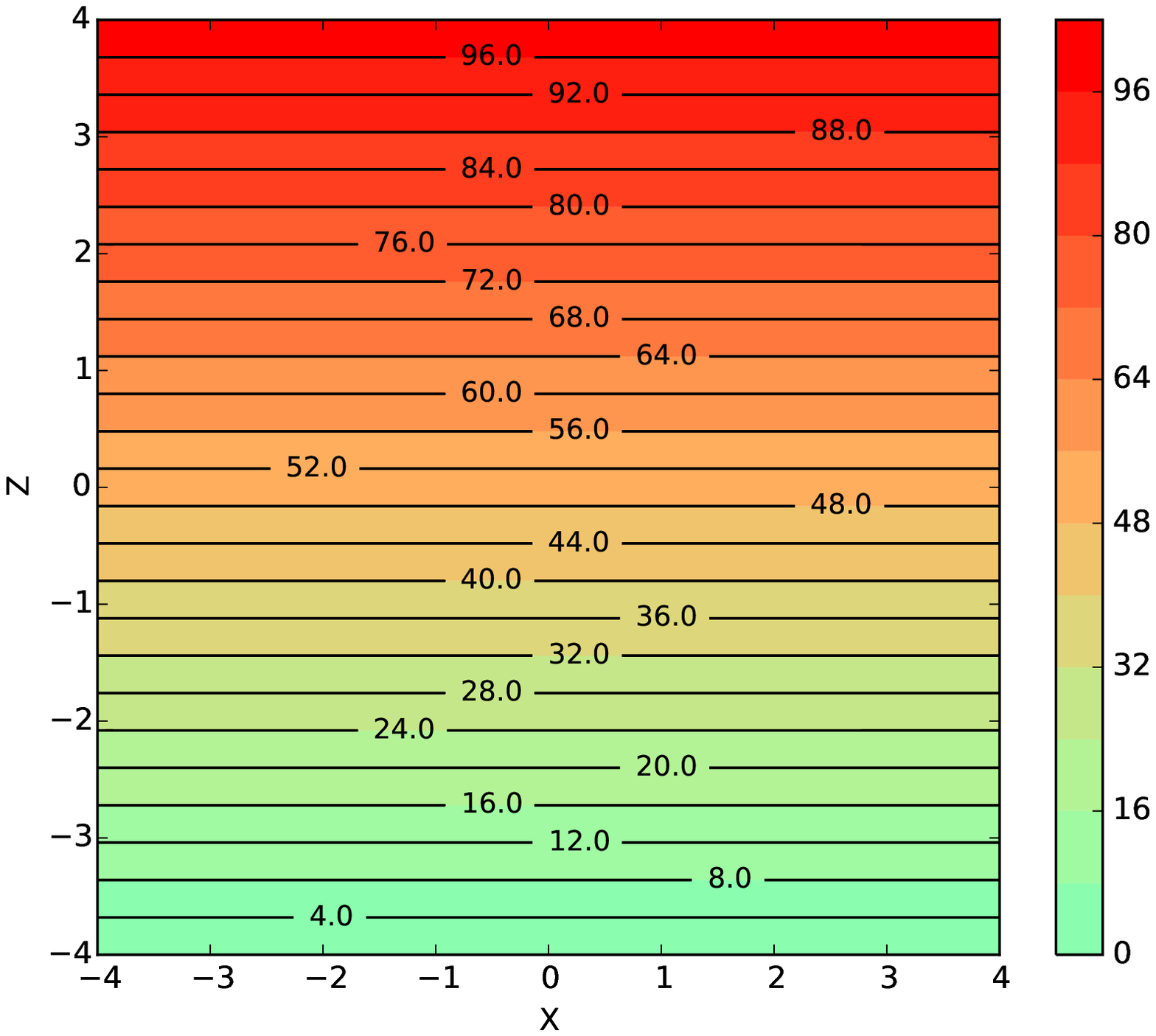}
\hspace{-3mm} \includegraphics[width=70mm, height=62mm]{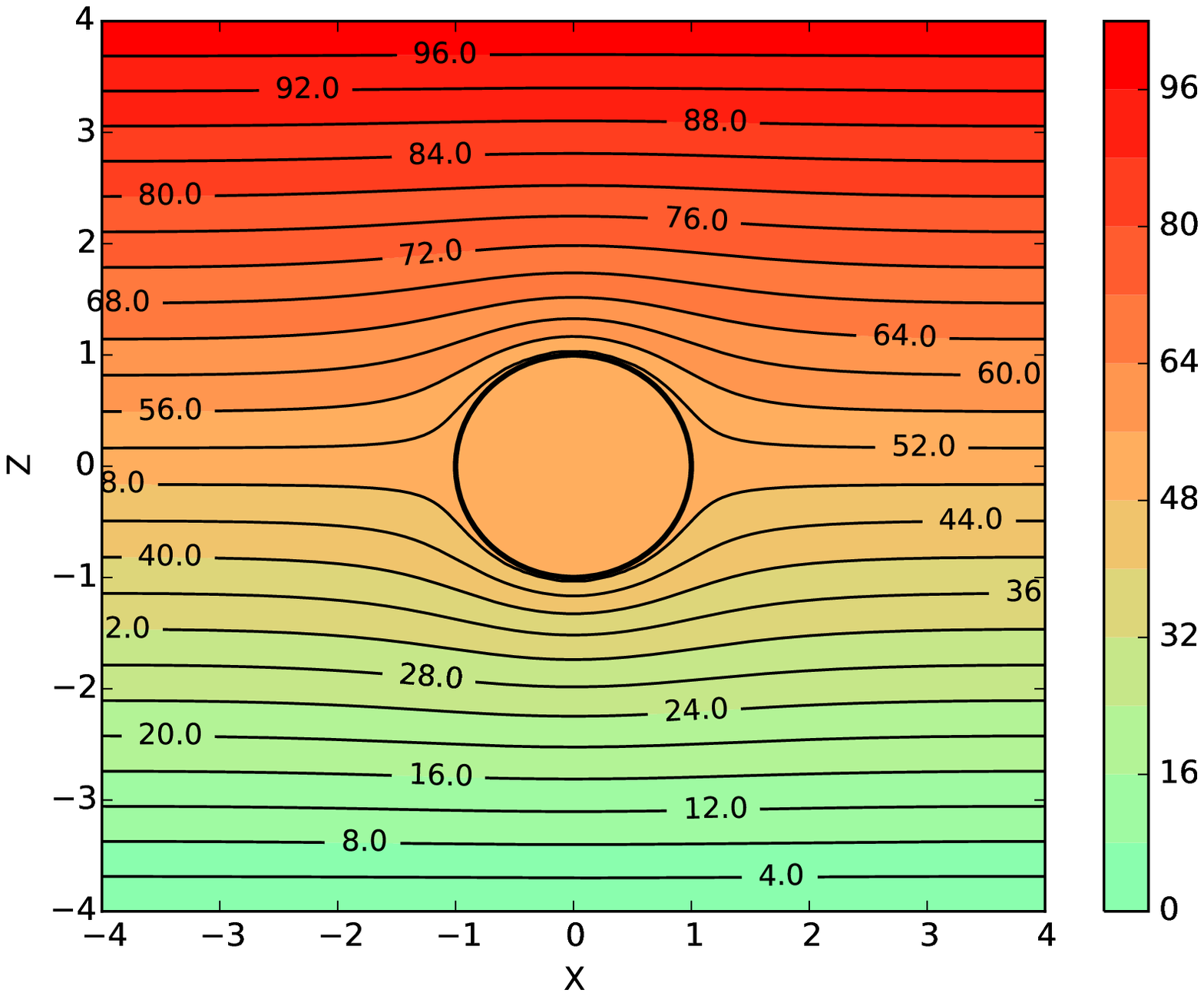}}

$\;$ \hspace{31mm} (a) \hspace{63mm} (b)
\caption{Equipotential lines (a) for the external field $U^e$ and (b) for an aluminum ball without a shell.}
\label{fig:2}
\end{figure}

In numerical experiments we assume that  domain $D$ and shell $\Omega$ are determined by the values 
$$
z_0 = 4 {\rm ~m},\; a = 1{\rm ~m}, \; b = 1.5{\rm ~m}, \; c = 4 {\rm ~m}.
$$ 
The role of the external field was played by the field $U^e$ in (\ref{eq7}) at $u_1 = 0$ V,
$u_2 = 100$ V which is characterized by straight equipotential lines $U^e={\rm const}$ oriented perpendicular to the $z$-axis (see Fig.~\ref{fig:2}(a)). In view of the axial symmetry of the direct boundary value problem, its solution in spherical coordinates is independent of the angle $\varphi$. With this in mind, an approximate solution was calculated in the cross section $D_2$ of three-dimensional domain $D$ by the plane $y = 0$.

Let us  briefly describe the cloaking scenario.
The field $U^e$, defined by formula (\ref{eq8}) and represented by isolines in Fig.~\ref{fig:2}(a) corresponds to
a homogeneous medium that fills entire domain $D$.
Placing an object with a different conductivity into this medium causes perturbation of the field $U^e$.
To demonstrate this effect we assume that $\Omega_i$ is an aluminum ball $r< a$ with the electric conductivity
$\sigma_i = 3.8 \times 10^7$ S/m while the rest of domain $D$ is filled with the stainless steel
($\sigma=1.45 \times 10^6$ S/m in $\Omega$ and $\Omega_e$). Solving direct problem (\ref{eq1})--(\ref{eq6}) 
corresponding to this situation we obtain electric potential $U$ with equipotential lines shown in Fig.~\ref{fig:2}(b).  
Perturbation of the field $U^e$ indicates the presence of body $(\Omega_i,\sigma_i)$ in $D$.
In order to suppress this perturbation we need to design a cloaking shell.

First, we solve control problems (\ref{eqJe}) and (\ref{eqJ}) for a single-layer anisotropic shell 
$(\Omega,\sigma_r,\sigma_\theta)$ assuming that the background material in $\Omega_i$ and $\Omega_e$ is stainless steel ($\sigma_b=1.45 \times 10^6$ S/m) while  
lower and upper bounds of control set $S$ in (\ref{eqS}) are determined by
\[
m_r=1.0 \times 10^2 \mbox{ S/m}, \;\;\; M_r=1.25 \times 10^5 \mbox{ S/m},
\]
\[
m_\theta=1.25 \times 10^5 \mbox{ S/m}, \;  \;\; M_\theta=3.8 \times 10^7 \mbox{ S/m}. 
\]
Numerical solution of problem (\ref{eqJe}) with the help of PSO method gives after 50 iterations the following results:
\[
\sigma_r^{opt}=61725 \mbox{ S/m}, \; \; \; \sigma_\theta^{opt}=1.77 \times 10^7 \mbox{ S/m},
\]
\[
J_e^{opt} \equiv J_e ({\bf s}^{opt}) =2.5 \times 10^{-5}, \; J_i ({\bf s}^{opt}) =1.1 \times 10^{-4}, \;
J ({\bf s}^{opt}) =6.8 \times 10^{-4}, \;\; {\bf s}^{opt} = (\sigma_r^{opt}, \sigma_\theta^{opt}).
\]
Let us note that the obtained solution $(\sigma_r^{opt}, \sigma_\theta^{opt})$ satisfies the admissibility 
condition $2 \sigma^{opt}_r \sigma^{opt}_\theta = \sigma_b^2 + \sigma_b \sigma^{opt}_r$ (see (\ref{AC})) with sufficient accuracy: $2 \sigma^{opt}_r \sigma^{opt}_\theta  / (\sigma_b^2 + \sigma_b \sigma^{opt}_r) \approx 0.997$.
This demonstrates the high accuracy of the proposed algorithm.

\begin{figure}[t!]
\centerline{
\includegraphics[width=72mm, height=65mm]{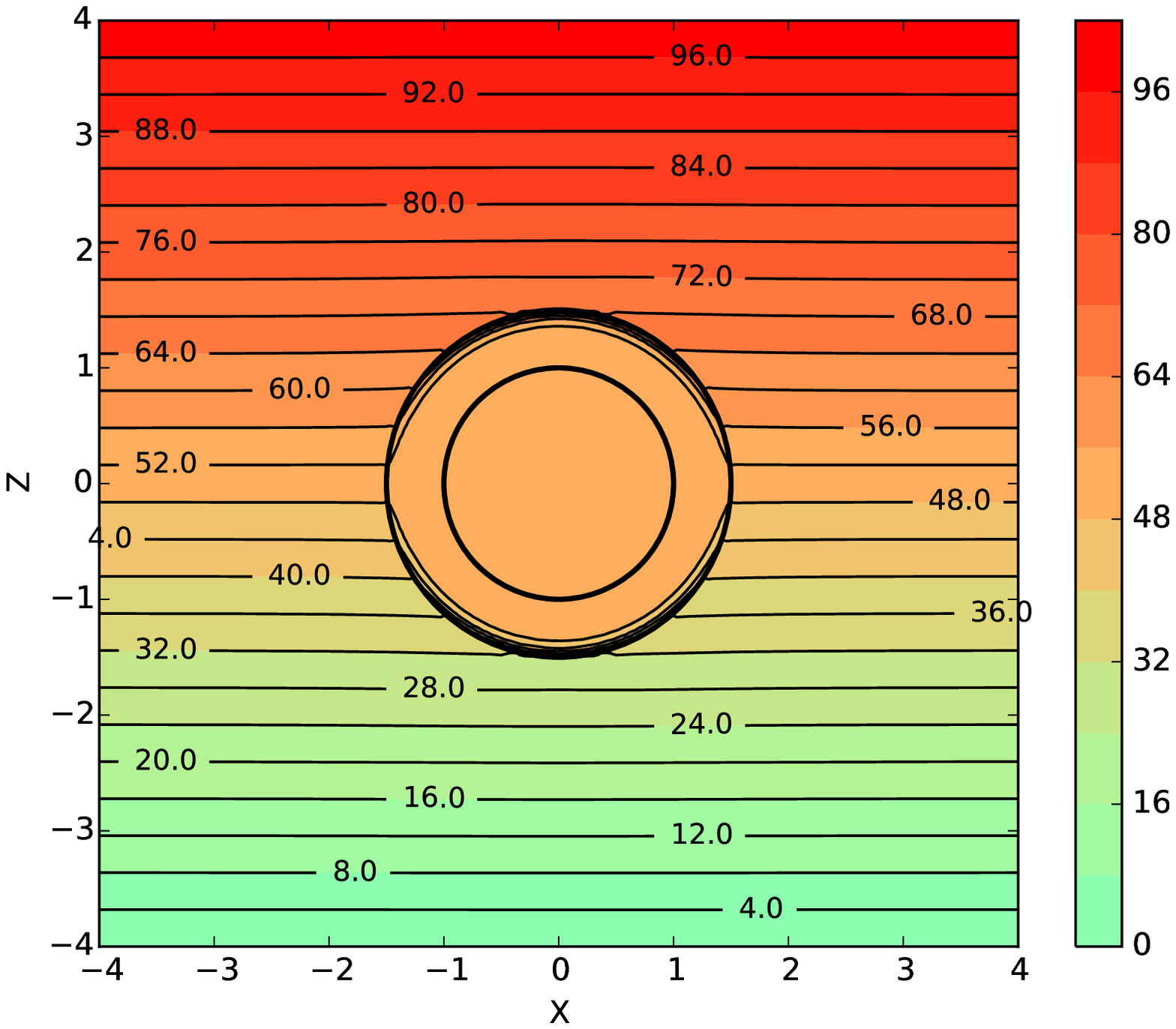} \hspace{-3mm} 
\includegraphics[width=72mm, height=65mm]{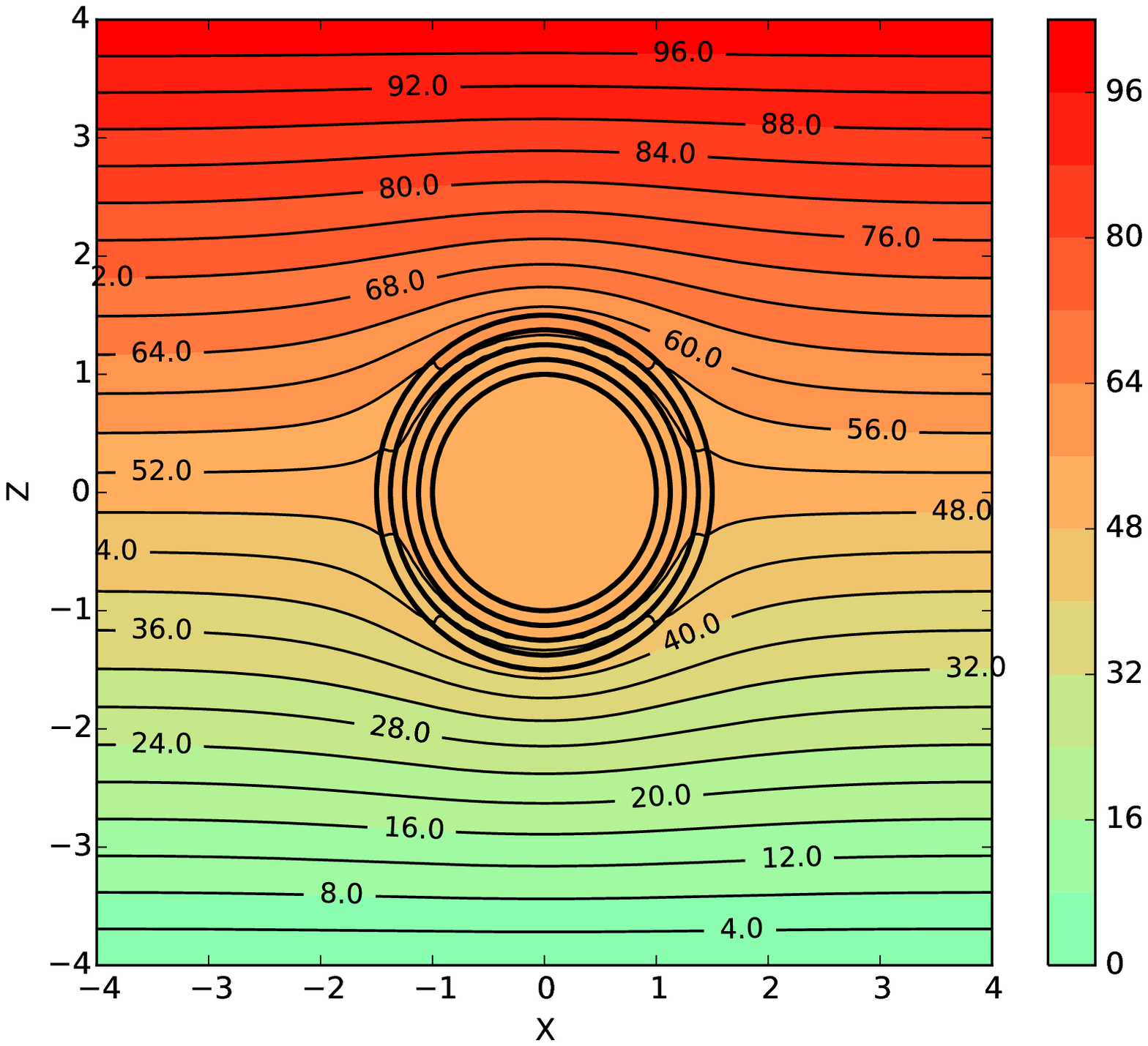}} 

$\;$ \hspace{31mm} (a) \hspace{63mm} (b)
\caption{Equipotential lines (a) for cloaking shell with a single anisotropic layer and (b) for unoptimized four-layer shell.}
\label{fig:3}
\end{figure}

If we solve the problem (\ref{eqJ}) then we obtain the following optimal values of control parameters and 
cost functionals:
\[
\sigma_r^{opt}=57562 \mbox{ S/m}, \;  \; \; \sigma_\theta^{opt}=1.89 \times 10^7 \mbox{ S/m}, 
\]
\[
J_e^{opt} \equiv J_e ({\bf s}^{opt}) =4.0 \times 10^{-5}, \; J_i ({\bf s}^{opt}) =5.5 \times 10^{-5}, \;
J ({\bf s}^{opt}) =2.5 \times 10^{-5}.
\]
Equipotential lines for corresponding field $U[{\bf s}^{opt}]$ are shown in Fig.~\ref{fig:3}(a).
It can be seen that electric field $U[{\bf s}^{opt}]$ in $\Omega_e$ 
is close to external field $U^e$ in Fig.~\ref{fig:2}(a)
and it creates the illusion of the absence of shell $\Omega$ for an external observer. 
It should be noted however that this high cloaking efficiency is achieved in both cases due to high anisotropy of the respective cloak $(\Omega,\sigma_r^{opt},\sigma_\theta^{opt})$.
It is characterized by the anisotropy parameter $s^{opt}=\sigma_\theta^{opt}/\sigma_r^{opt}$
which is equal to 286 in the first case and 328 in the second case.

As mentioned above, technical realization of highly anisotropic cloaking shells is associated with great difficulties because appropriate materials are not found in nature and should be created using metamaterials approach.
That is why our second group of tests deals with designing multilayer shells with isotropic layers.
Let an even integer $M$ denote the number of layers of our shell and $\sigma_j$ be electric conductivity of the $j$th layer, $j=1,\ldots,M$.
Setting ${\bf s}=(\sigma_1, \sigma_2, \ldots, \sigma_M)$  we  define the control set $S$ as follows
\begin{equation}
\label{eqK} 
S=\{ {\bf s}:   \sigma_{min} \le \sigma_j \le \sigma_{max}, j=1,\dots,M  \}.  
\end{equation}
Here given positive constants $\sigma_{min}$ and $\sigma_{max}$ have the sense of lower and upper bounds of $S$.


We begin our analysis of the cloaking properties of layered isotropic shells with consideration 
of the alternating design strategy. 
For two materials with conductivities $\sigma_{min}$ and $\sigma_{max}$ it responds to a scenario in which materials are distributed in layers in such a way that conductivities $\sigma_j=\sigma_j^{alt}$ of the separate layers $\Omega_j$ satisfy 
\begin{equation}
\label{Kalt}
\sigma_1^{alt}=\sigma_3^{alt}=\ldots=\sigma_{M-1}^{alt}=\sigma_{min}, \;\;\;
\sigma_2^{alt}=\sigma_4^{alt}=\ldots=\sigma_M^{alt} = \sigma_{max}. 
\end{equation}
Let us first choose the following values: 
$\sigma_{min} = 1.2 \times 10^5$ S/m (corresponds to graphite), $\sigma_{max} = 3.8 \times 10^7$ S/m 
(corresponds to aluminum) and $\sigma_b = 1.45 \times 10^6$ S/m (stainless steel). Since 
conductivities $\sigma_j$ of all layers $\Omega_j$  are known we are able to  determine the shielding or cloaking performance of the corresponding shell $(\Omega, {\bf s}^{alt})$ by calculating the values $J_i({\bf s}^{alt})$ or 
$J_e({\bf s}^{alt})$ and $J ({\bf s}^{alt})$ for the conductivity vector 
${\bf s}^{alt}=(\sigma_{min}, \sigma_{max}, \ldots, \sigma_{min}, \sigma_{max})$.
These values describing the corresponding mean square errors of fulfilling
the first or second condition or both conditions in (\ref{IP2})
together with the value $\sigma_M^{alt} = \sigma_{max} = 3.8 \times 10^7$ S/m are presented for 
six different shells corresponding to $M=2$, $4$, $6$, $8$, $10$ and $12$  in Table~1. 
Isolines of corresponding  field $U^{alt} \equiv U [{\bf s}^{alt}]$ for a six-layer cloaking shell ($M=6$) are shown in Figure~\ref{fig:3}(b).
It is seen from Table~1 that the values $J_i({\bf s}^{alt})$, $J_e({\bf s}^{alt})$ and $J({\bf s}^{alt})$ decrease with increasing $M$ but very slowly. In particular, $J({\bf s}^{alt})=6.6 \times 10^{-2}$ for $M=12$. 
Besides, the isolines outside the shell are curved. 
These results demonstrate low shielding and cloaking performances of the cloak $(\Omega,{\bf s}^{alt})$ based on alternating design for the given couple $(\sigma_{\rm min}, \sigma_{\rm max})$.

\begin{table}[htb]
\caption{Numerical results for nonoptimized multilayer shells 
($\sigma_{min} = 1.2 \times 10^5$ S/m,  $\sigma_{max} = 3.8 \times 10^7$ S/m).}
\begin{center}
\begin{tabular}{ccccc}
\hline
$M$   & $\sigma_M$ [S/m]  & $J_i({\bf s}^{alt})$   & $J_e({\bf s}^{alt})$ & $J({\bf s}^{alt})$ \\
\hline
  2 \rule{0cm}{4mm}    & $3.8 \times 10^7$ & $1.2 \times 10^{-1}$ & $6.2 \times 10^{-2}$    & $1.8 \times 10^{-1}$\\
  4    & $3.8 \times 10^7$  & $3.8 \times 10^{-2}$ & $4.9 \times 10^{-2}$ & $8.7 \times 10^{-2}$ \\
  6    & $3.8 \times 10^7$  & $2.3 \times 10^{-2}$ & $6.3 \times 10^{-2}$ & $8.6 \times 10^{-2}$ \\
  8    & $3.8 \times 10^7$  & $1.8 \times 10^{-2}$ & $5.6 \times 10^{-2}$ & $7.4 \times 10^{-2}$  \\
  10   & $3.8 \times 10^7$  & $1.7 \times 10^{-2}$ & $5.2 \times 10^{-2}$ & $6.9 \times 10^{-2}$ \\
  12   & $3.8 \times 10^7$  & $1.6 \times 10^{-2}$ & $5.0 \times 10^{-2}$ & $6.6 \times 10^{-2}$ \\
\hline
\end{tabular}
\end{center}
\end{table}

Now we expand the control set $S$ in (\ref{eqK}) using the other two couples 
$(\sigma_{\rm min}, \sigma_{\rm max})$ namely: $\sigma_{\rm min} =1.0 \times 10^4$ S/m (corresponds to carbon),
$\sigma_{\rm max}= 3.8 \times 10^7$ S/m (corresponds to aluminum) and $\sigma_{\rm min} =1.0 \times 10^4$ S/m while
$\sigma_{\rm max}= 5.9 \times 10^7$ S/m (corresponds to copper), respectively.
The conductivities $\sigma_j^{alt}$ of separate layers $\Omega_j$ corresponding to these couples 
are given in (\ref{Kalt}), while  $\sigma_M^{alt}$ and the corresponding values 
$J_i({\bf s}^{alt})$, $J_e({\bf s}^{alt})$ and $J ({\bf s}^{alt})$
of all cost functionals  are presented in Tables~2 and 3.
Analysis of Tables 1--3 shows that increasing the number of layers $M$ and the contrast $\sigma_{max} / \sigma_{min}$ 
{ decreases the value} $J_i({\bf s}^{alt})$ to $4.3 \times 10^{-7}$ and therefore substantially { increases} the shielding performance of the respective shell $(\Omega,{\bf s}^{alt})$. However, increasing $M$ does not lead to a decrease in $J({\bf s}^{alt})$.
Therefore in all three cases the cloaking performance of the corresponding shell $(\Omega,{\bf s}^{alt})$ remains low.

\begin{table}[htb]
\caption{Numerical results for nonoptimized multilayer shells 
($\sigma_{min} = 1.0 \times 10^4$ S/m,  $\sigma_{max} = 3.8 \times 10^7$ S/m).}
\begin{center}
\begin{tabular}{ccccc}
\hline
$M$   & $\sigma_M$ [S/m] & $J_i({\bf s}^{alt})$   & $J_e({\bf s}^{alt})$  & $J({\bf s}^{alt})$  \\
\hline
  2 \rule{0cm}{4mm}    & $3.8 \times 10^7$ & $1.4 \times 10^{-2}$ & $6.2 \times 10^{-2}$  & $7.6 \times 10^{-2}$ \\
  4    & $3.8 \times 10^7$ & $5.8 \times 10^{-4}$ & $4.6 \times 10^{-2}$  & $4.6 \times 10^{-2}$ \\
  6    & $3.8 \times 10^7$ & $6.7 \times 10^{-5}$ & $5.2 \times 10^{-2}$  & $5.2 \times 10^{-2}$ \\
  8    & $3.8 \times 10^7$ & $1.4 \times 10^{-5}$ & $3.9 \times 10^{-2}$  & $3.9 \times 10^{-2}$ \\
  10   & $3.8 \times 10^7$ & $7.0 \times 10^{-6}$ & $2.9 \times 10^{-2}$  & $2.9 \times 10^{-2}$ \\
  12   & $3.8 \times 10^7$ & $3.1 \times 10^{-6}$ & $2.3 \times 10^{-2}$  & $2.3 \times 10^{-2}$ \\
\hline
\end{tabular}
\end{center}
\end{table}

\begin{table}[htb]
\caption{Numerical results for nonoptimized multilayer shells 
($\sigma_{min} = 1.0 \times 10^4$ S/m,  $\sigma_{max} = 5.9 \times 10^7$ S/m).}
\begin{center}
\begin{tabular}{ccccc}
\hline
$M$   & $\sigma_M$ [S/m] & $J_i({\bf s}^{alt})$   & $J_e({\bf s}^{alt})$  & $J({\bf s}^{alt})$  \\
\hline
  2 \rule{0cm}{4mm}    & $5.9 \times 10^7$ & $9.8 \times 10^{-3}$ & $6.9 \times 10^{-2}$   & $7.9 \times 10^{-2}$\\
  4    & $5.9 \times 10^7$  & $2.7 \times 10^{-4}$ & $5.7 \times 10^{-2}$ & $5.7 \times 10^{-2}$\\
  6    & $5.9 \times 10^7$  & $2.2 \times 10^{-5}$ & $7.1 \times 10^{-2}$ & $7.1 \times 10^{-2}$ \\
  8    & $5.9 \times 10^7$  & $3.4 \times 10^{-6}$ & $5.9 \times 10^{-2}$ & $5.9 \times 10^{-2}$ \\
  10   & $5.9 \times 10^7$  & $1.2 \times 10^{-6}$ & $4.9 \times 10^{-2}$ & $4.9 \times 10^{-2}$ \\
  12   & $5.9 \times 10^7$  & $4.3 \times 10^{-7}$ & $4.2 \times 10^{-2}$ & $4.2 \times 10^{-2}$ \\
\hline
\end{tabular}
\end{center}
\end{table}


In order to obtain shielding or cloaking shells with substantially higher performances, 
the optimization method should be applied to the design of these shells. 
The PSO applied to all three problems (\ref{eqJi}),  (\ref{eqJe}) and (\ref{eqJ}) resulted in two surprising facts.
The first  discovered using the PSO algorithm concerns the solution of shielding problem
(\ref{eqJi}) for all three previously chosen couples $(\sigma_{min}, \sigma_{max})$. As it turned out, in all three cases the optimal values of conductivities $\sigma^{opt}_j$ of the separate layers coincide exactly with the values of 
$\sigma^{alt}_j$ in (\ref{Kalt}), corresponding to the strategy of alternating design.
Corresponding optimal values $J_i^{opt} \equiv J_i({\bf s}^{alt})$  of the functional  
$J_i({\bf s})$ describing the respective mean square errors of the shielding problem solution
 are given in the  third column of  Tables~1, 2 and 3.

On the one hand, this fact means that for the shielding problem an analogue of the so-called bang-bang principle is valid. According to this principle (see, for example, \cite{Alpha99}), 
the solution to the control problem under study  takes values at the certain boundary points of the control set.
We emphasize that the use of this property allows us to formulate simple design rules for designing highly efficient shielding devices. The rules consist simply  in choosing the conductivities $\sigma_{min}$ and $\sigma_{max}$ of two different natural materials with high contrast  
$\sigma_{max} / \sigma_{min}$ as the lower  and the upper  bounds of the set $S$ in (\ref{eqK}).
In particular, choosing $\sigma_{min} = 1.0 \times 10^4$ S/m, $\sigma_{max} = 5.9 \times 10^7$ S/m, we obtained (see above) the shielding  shell $(\Omega, {\bf s}^{opt}) = (\Omega, {\bf s}^{alt})$ with a minimum mean square error
$J_i^{opt} = J_i ({\bf s}^{alt})$ equal to $4.3 \times 10^{-7}$ for $M = 12$, which corresponds to the highest shielding performance.

The second surprising fact that we discovered concerns the solution of the design problem of cloaking shells for all three different choices of couples  $(\sigma_{min}, \sigma_{max})$. As  our optimization analysis showed, the optimal values $\sigma^{opt}_j$ obtained using the PSO method satisfy the condition $\sigma^{opt}_j =\sigma^{alt}_j$ for all $j$  except $j = M$. In other words, in this case, instead of (\ref{Kalt}), the relations 
\begin{equation}
\label{Salt2} 
\sigma_1^{opt}=\sigma_3^{opt}=\ldots=\sigma_{M-1}^{opt}= \sigma_{min}, \;\;\;
\sigma_2^{opt}=\sigma_4^{opt}=\ldots=\sigma_{M-2}^{opt}=\sigma_{max}, \;\;\;
\sigma_{min} \le \sigma_{M}^{opt} \le \sigma_{max}
\end{equation}
hold. 
As for the last value $\sigma^{opt}_M$, then for each $M$, $\sigma^{opt}_M$ takes some intermediate value 
between $\sigma_{min}$ and $\sigma_{max}$ depending on the parameters of the cloaking problem under study. 

The values of $\sigma^{opt}_M$ and the corresponding optimal values
$J^{opt} = J ({\bf s}^{opt})$, $J_i ({\bf s}^{opt})$, $J_e ({\bf s}^{opt})$,
where $ {\bf s}^{opt} = (\sigma^{opt}_1, \sigma^{opt}_2, \ldots, \sigma^{opt}_M)$, for various $M = 2,4 ,\ldots, 12$ are presented in Tables 4--6 for three different ways of 
choosing couples $(\sigma_{min}, \sigma_{max})$.

\begin{table}[th]
\caption{Numerical results for optimized multilayer shells 
($\sigma_{min} = 1.2 \times 10^5$ S/m,  $\sigma_{max} = 3.8 \times 10^7$ S/m).}
\begin{center}
\begin{tabular}{ccccc}
\hline
$M$   & $\sigma^{opt}_M$ [S/m] & $J_i({\bf s}^{opt})$   & $J_e({\bf s}^{opt})$  & $J({\bf s}^{opt})$ \\
\hline
  2 \rule{0cm}{4mm}    & $3.80 \times 10^7$  & $1.2 \times 10^{-1}$ & $6.2 \times 10^{-2}$  & $1.8 \times 10^{-1}$ \\
  4    & $3.22 \times 10^6$  & $8.5 \times 10^{-2}$ & $2.6 \times 10^{-5}$ & $8.5 \times 10^{-2}$ \\
  6    & $1.21 \times 10^6$  & $4.6 \times 10^{-2}$ & $7.6 \times 10^{-5}$ & $4.6 \times 10^{-2}$ \\
  8    & $4.59 \times 10^5$  & $2.8 \times 10^{-2}$ & $8.5 \times 10^{-5}$ & $2.8 \times 10^{-2}$ \\
  10   & $1.20 \times 10^5$  & $1.9 \times 10^{-2}$ & $7.4 \times 10^{-3}$ & $2.6 \times 10^{-2}$ \\
  12   & $1.20 \times 10^5$  & $1.8 \times 10^{-2}$ & $1.5 \times 10^{-3}$ & $2.0 \times 10^{-2}$ \\
\hline
\end{tabular}
\end{center}
\end{table}

\begin{table}[htb]
\caption{Numerical results for optimized multilayer shells 
($\sigma_{min} = 1.0 \times 10^4$ S/m,  $\sigma_{max} = 3.8 \times 10^7$ S/m).}
\begin{center}
\begin{tabular}{ccccc}
\hline
$M$   & $\sigma_M$ [S/m] & $J_i({\bf s}^{opt})$   & $J_e({\bf s}^{opt})$  & $J({\bf s}^{opt})$ \\
\hline
  2 \rule{0cm}{4mm}   & $4.36 \times 10^7$ & $4.9 \times 10^{-2}$ & $3.4 \times 10^{-5}$ & $4.9 \times 10^{-2}$ \\
  4    & $8.20 \times 10^6$  & $1.2 \times 10^{-3}$ & $1.6 \times 10^{-5}$ & $1.2 \times 10^{-3}$\\
  6    & $1.17 \times 10^7$  & $1.4 \times 10^{-4}$ & $1.4 \times 10^{-5}$ & $1.5 \times 10^{-4}$ \\
  8    & $1.50 \times 10^7$  & $2.0 \times 10^{-5}$ & $1.3 \times 10^{-5}$ & $3.3 \times 10^{-5}$ \\
  10   & $1.81 \times 10^7$  & $7.1 \times 10^{-6}$ & $9.1 \times 10^{-6}$ & $1.6 \times 10^{-5}$ \\
  12   & $2.09 \times 10^7$  & $3.8 \times 10^{-6}$ & $9.1 \times 10^{-6}$ & $1.3 \times 10^{-5}$ \\
\hline
\end{tabular}
\end{center}
\end{table}

\begin{table}[htb]
\caption{Numerical results for optimized multilayer shells 
($\sigma_{min} = 1.0 \times 10^4$ S/m,  $\sigma_{max} = 5.9 \times 10^7$ S/m).}
\begin{center}
\begin{tabular}{ccccc}
\hline
$M$   & $\sigma_M$ [S/m] & $J_i({\bf s}^{opt})$   & $J_e({\bf s}^{opt})$  & $J({\bf s}^{opt})$ \\
\hline
  2 \rule{0cm}{4mm} & $4.36 \times 10^7$ & $4.9 \times 10^{-2}$ & $3.4 \times 10^{-5}$ & $4.9 \times 10^{-2}$\\
  4    & $8.19 \times 10^6$ & $7.9 \times 10^{-4}$ & $1.6 \times 10^{-5}$  & $8.1 \times 10^{-4}$\\
  6    & $1.16 \times 10^7$ & $5.9 \times 10^{-5}$ & $2.4 \times 10^{-5}$  & $7.3 \times 10^{-5}$\\
  8    & $1.50 \times 10^7$ & $6.0 \times 10^{-6}$ & $1.3 \times 10^{-5}$  & $1.9 \times 10^{-5}$\\
  10   & $1.79 \times 10^7$ & $2.4 \times 10^{-6}$ & $1.6 \times 10^{-5}$  & $1.8 \times 10^{-5}$ \\
  12   & $2.06 \times 10^7$ & $6.3 \times 10^{-7}$ & $9.1 \times 10^{-6}$  & $9.7 \times 10^{-6}$ \\
\hline
\end{tabular}
\end{center}
\end{table}

\begin{figure}[h!]
\centerline{
\includegraphics[width=71mm, height=64mm]{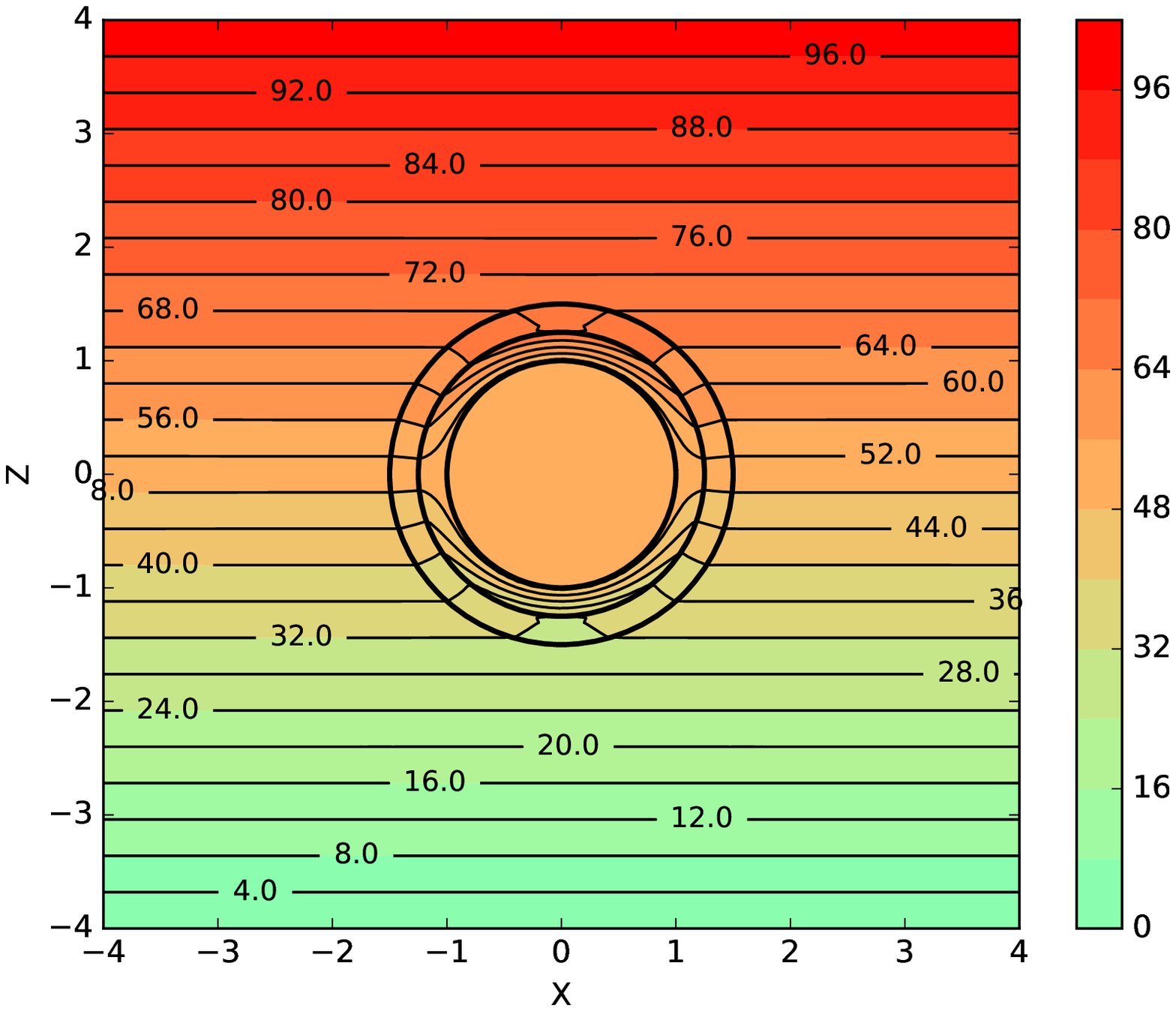} 
\hspace{-3mm} 
\includegraphics[width=70mm, height=65mm]{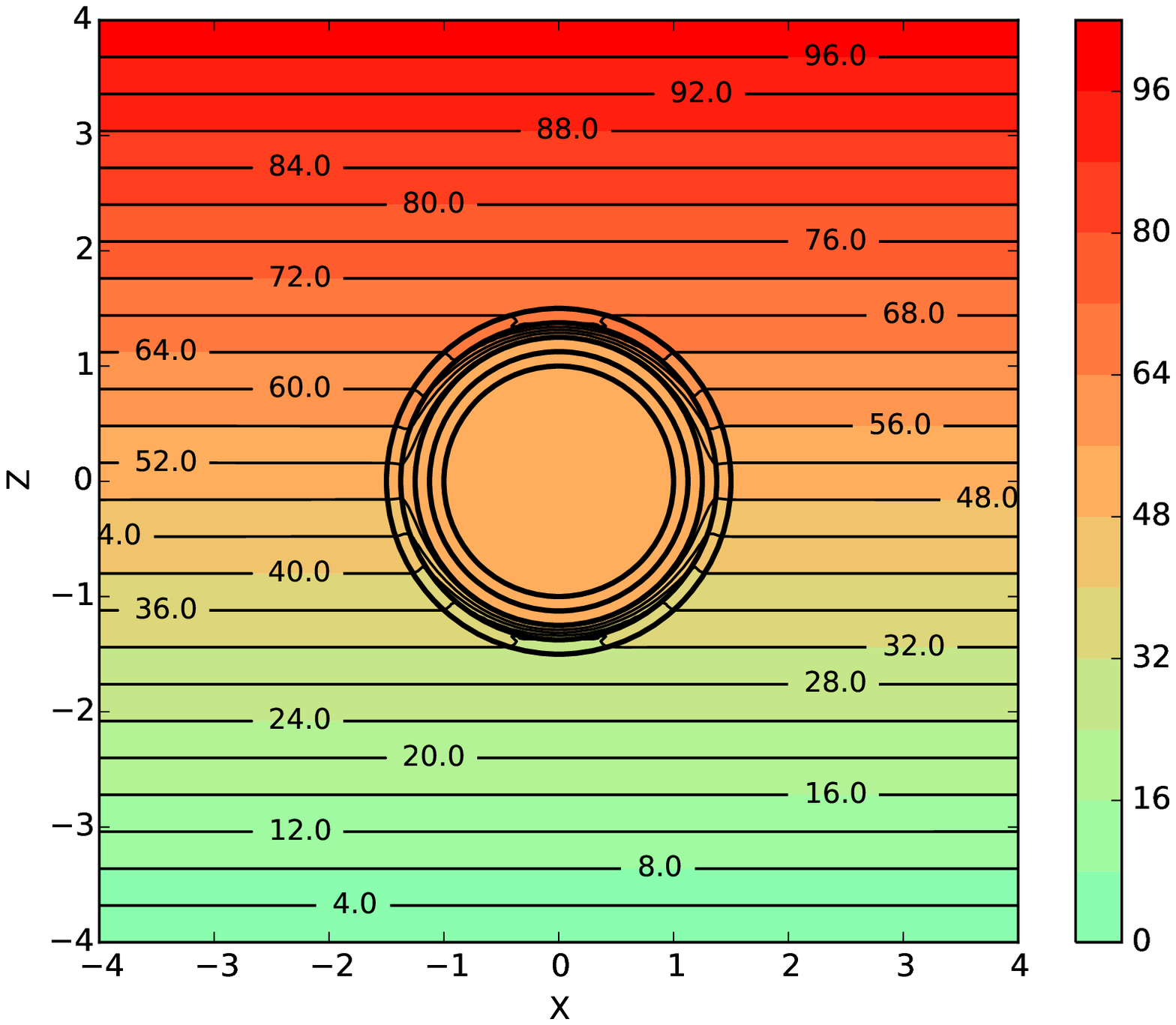}}

$\;$ \hspace{32mm} (a) \hspace{63mm} (b)
\caption{Equipotential lines for shell with (a) two and (b) four isotropic layers.}
\label{fig:4}
\end{figure}

Analysis of Tables~4--6 shows that the last choice of the bounds $\sigma_{min}$ and  $\sigma_{max}$
gives the least optimal values $J^{opt}$ of functional $J$ (in particular, $J^{opt} = 9.7 \times 10^{-6}$) at $M=12$ which corresponds to the highest cloaking
performance of the shell $(\Omega,{\bf s}^{opt})$.
Equipotential lines of the corresponding field $U[{\bf s}^{opt}]$ for  two- and four-layer shells
are shown in Fig.~\ref{fig:4}(a) and Fig.~\ref{fig:4}(b), respectively.
In both cases the electric potential $U[{\bf s}^{opt}]$ in $\Omega_e$ is close to the external field $U^e$
shown in Fig.~\ref{fig:2}(a).
Thus to obtain a shell with the highest cloaking performance it is sufficient to choose the smallest possible value
of the lower bound $\sigma_{min}$ and the largest value of the upper bound  $\sigma_{min}$.

The fact that optimal values of all control parameters except the last  $\sigma_M^{opt}$
are equal to the lower or upper bounds of control set $S$ means that
the design of a multilayer cloak by solving $M$-dimensional control problem (\ref{eqJ})
can be reduced to a one-parameter minimization problem with respect to $\sigma_M$. 
On the one hand, this greatly simplifies the solution of problem (\ref{eqJ}). 
On the other hand, this means that the optimal cloaking shell $(\Omega, {\bf s}^{opt})$ in its structure consists of three different (in general) materials. The first $M-1$ 
layers of the optimal shell are filled (as in the case of shielding shells) with alternating materials corresponding to the couple $(\sigma_{min}, \sigma_{max})$, while the last layer is filled with the material corresponding to the optimal value of $\sigma^{opt}_M$.

If the found material corresponding to $\sigma^{opt}_M$  belongs to the class of natural materials, then the result of our optimization design is an easily manufactured $M$-layered isotropic shell where all layers are made of natural materials. For example, the value $\sigma^{opt}_{10} = 1.79 \times 10^7$ S/m given in Table~6  at $M = 10$ describes the conductivity of tungsten. 
This means that applying the proposed optimization design with 
$\sigma_{min} = 1.0 \times 10^4$ S/m, $\sigma_{max} = 5.9 \times 10^7$ S/m
allows us to obtain a ten-layer cloaking shell, the first nine layers of which are filled with alternating carbon and copper,  while the last layer is filled with tungsten. 
We emphasize that { the total mean square error}
$J^{opt} \equiv J ({\bf s}^{opt})$ for this shell is equal to $1.8 \times 10^{-5}$ (see Table~6), which corresponds to the very high cloaking performance.

If the optimal value $\sigma_M^{opt}$ does not correspond to any natural material, then one may achieve an easily manufactured cloaking shell by selecting a  value  $\tilde \sigma_M^{opt}$ in the vicinity of $\sigma_M^{opt}$   corresponding to some natural material or alloy. For example, instead of 
$\sigma_{6}^{opt} = 1.16 \times 10^7$ S/m presented in Table 6 for $M=6$, one 
can use a close value $\tilde \sigma_{6}^{opt} = 1.15 \times 10^7$ S/m
corresponding to nickel. Setting $\tilde {\bf s}^{opt} = (\sigma_1^{opt}, \ldots, \sigma_{M-1}^{opt}, \tilde \sigma_M^{opt})$ we obtain the cloaking shell  $(\Omega, \tilde {\bf s}^{opt})$ with the error 
$J(\tilde {\bf s}^{opt}) = 5.5 \times 10^{-4}$, which again corresponds to a high cloaking performance.

An alternative way to design easily manufactured high-performance shells is to use geometric parameters as additional controls such as the widths of several or all layers.

\section{Conclusion}

In this paper we have studied inverse problems for a 3D model of DC conduction (\ref{eq1})--(\ref{eq6}) associated with designing spherical devices serving for manipulation of DC electric fields. 
A particular attention has been paid to inverse problems that arise in the design of axisymmetric DC 
shielding or cloaking layered shells. Using an optimization approach these inverse problems have been reduced to corresponding control problems in which the electric conductivities play the role of control parameters. The developed numerical algorithm employs particle swarm optimization. Numerous results of computational experiments and optimization analysis have shown that high performances of the shells can be achieved by using either a highly anisotropic single-layer shells or a multilayer shells with isotropic layers. In the latter case it turned out that with a certain choice of a control set for optimal solutions an analogue of the bang-bang principle holds true. By virtue of this principle optimal solutions have a simple structure  coinciding with that of the solution that corresponds to the alternating design for the shielding problem or differs only by the conductivity of the last layer in the cloaking problem. Based on the established structure of optimal solutions simple rules for the design of highly efficient shielding or cloaking devices have been formulated. The simulation results have confirmed excellent cloaking or shielding performances of the shells designed on the basis of our method. In addition we have demonstrated that these shells can be easily manufactured. Finally, we note that the proposed method is not limited to the DC electric fields and can be extended to the thermal, magnetic and other physical fields and to designing concentrators, inverters, and other functional devices for manipulating physical fields.

\section*{Acknowledgments}
The first author was supported by the Russian Foundation for Basic Research (project no. 16-01-00365-a) 
and the second author by the Federal Agency for Scientific Organizations in the framework of the State Task Program  (subject no. 0263-2018-0001).

\bibliographystyle{IEEEtran}

\end{document}